\documentclass[11pt,a4paper]{article}   
\usepackage{amsmath}
\usepackage{amssymb}
\usepackage{mathrsfs}

\usepackage{rotating}

\usepackage[round]{natbib}

\begin{document}

\title{Surge motion of an ice floe in waves: comparison of theoretical and experimental models}

\author{Michael H. MEYLAN$^1$ Lucas J. YIEW$^2$ Luke G. BENNETTS$^2$
\\
Benjamin J. FRENCH$^3$ Giles A. THOMAS$^{3}$ 
\\
{\footnotesize
$^1$School of Mathematical and Physical Science, The University of Newcastle, NSW 2308, Australia}
\\
{\footnotesize
$^2$School of Mathematical Sciences, University of Adelaide, SA 5005, Australia}
\\
{\footnotesize
$^3$National Centre for Maritime Engineering and Hydrodynamics, Australian Maritime College, Launceston 7250, Australia}
}

\date{\today}
\maketitle

\abstract{
A theoretical model and an experimental model of surge motions of an ice floe due to regular waves are presented.
The theoretical model is a modified version of Morrison's equation, valid for small floating bodies.
The experimental model is implemented in a wave basin at scale 1:100, using a thin plastic disk to model the floe.
The processed experimental data displays a regime change in surge amplitude when the incident wavelength is approximately twice the floe diameter.
It is shown that the theoretical model is accurate in the large wavelength regime, but highly inaccurate for the small wavelength regime. 
}

\section{Introduction}

Ocean surface waves that penetrate into the ice-covered ocean force the ice floes there to surge back and forth.
The surge motion is most significant over the first O(10\,km) in from the ice edge, where
the waves have not been damped significantly by the ice cover, and where
the average floe size is relatively small, due to wave-induced ice fracture.

Surge motions cause  floes to collide 
if the amplitude of the surge motion is on the order of the floe separations, 
and if motions of adjacent floes are sufficiently out of phase. 
Collisions transfer momentum through the ice cover, 
thus affecting its large-scale rheology \citep{shenetal87,feltham05}.
Floe edges are also eroded by collisions \citep{mckenna90}.
Further, collisions can result in floes rafting on top of one another.

In order to model wave-induced floe-floe interactions accurately, 
it is first necessary to model wave-induced surge motions accurately.
Here we present a theoretical model 
and laboratory experimental model of surge motions of an ice floe, in which the floe is modelled as a thin floating disk.
The theoretical model is based on a modified version of Morrison's equation, which includes a force due to the slope of the wave field.
Morrison's equation is a semi-empirical model, which assumes that the body (here disk/floe) does not affect the surrounding wave field, i.e. it is small in relation to the wavelength. 
We use the experimental model to validate the theoretical model, and to establish its range of validity for a thin floating disk.

The modified version of Morrison's equation was developed by \citet{rumer_etal79} and \citet{marchenko99} for the ice floe problem.
We refer to this model as the Rumer model.  
\cite{shen_ackley91} analysed the Rumer model as part of their investigation into the drift of pancake ice.
\cite{shen_zhong01} derived an analytic approximate solution for the Rumer model.
\citet{grotmaack_meylan06}  proved that all solutions of the Rumer model tend to the same steady-state solution
(for moderate wave steepnesses). 

A series of  laboratory experiments, closely related to ours, were reported by \cite{huang_etal11} and \citet{huang_law13}. 
The experiments were designed to investigate drift of a thin floating body in waves. 
However, experimental measurement of drift in a wave tank is extremely challenging because 
Stokes drift induces a large scale circulatory motion in the tank. 
This can corrupt the drift and lead to anomalous results with drift velocities significantly
exceeding Stokes drift. 

In the following sections we present the Rumer model, and describe the experimental campaign and data analysis performed. 
A comparison of surge motions predicted by the theoretical and experimental models is then presented.
We find that the Rumer model is accurate when wavelengths are greater than twice the floe diameter.
Heave and pitch motions of the disk measured during the experiments are also presented.

\section{Theoretical Model}

\citet{rumer_etal79} and \citet{marchenko99} developed models of horizontal motions of an ice floe due to waves, 
assuming that the floe is much smaller than the wavelength. 
We use a combination of the models of \citet{rumer_etal79} and \citet{marchenko99}, derived by \cite{grotmaack_meylan06}, which assumes low wave steepness.  
We restrict our investigation here to monochromatic waves only. 


\subsection{Wave Forcing on Small Floating Bodies}

The force acting on a small body (in relation to the wavelength) is decomposed into two components:
\begin{enumerate}
\item
a sliding force due to gravity; and
\item
a drag force between the water and the body.
\end{enumerate}


\subsubsection{Gravity}

The sliding force due to gravity, $F_g$, is given by
\begin{equation} \label{eq:sliding}
F_g = 
-m g 
\frac {\partial \eta} {\partial x},
\end{equation}
where $\eta$ is the profile of the water surface, $x(t)$ is the horizontal location of the body, and $g\approx 9.81$\,ms$^{-2}$ is gravitational acceleration.

For a monochromatic wave, of height $H$, wavenumber $k$, and frequency $f$, the wave profile at time $t$ is
\begin{align*}
\eta(x,t) &
 = 
\frac H 2
\sin( k x - 2\pi f t )
\end{align*}
The slope of the surface profile is therefore
\begin{equation} \label{eq:detadt}
\frac {\partial \eta} {\partial x} = 
\frac {kH} {2}
\cos( k x - 2\pi f t ).
\end{equation}

\subsubsection{Drag}

Drag force, $F_d$, is caused by friction between the body and the water. 
It is expressed as
\begin{equation} \label{eq:drag}
\begin{aligned}
F_d = & \rho C_d W  \left| V_w - V \right| (V_w - V).
\end{aligned}
\end{equation}
Here $\rho$ is the mass density of the fluid, $W$ is the reference (wetted surface) area, $C_d$ is the drag coefficient, and $V_w$ and $V$ are horizontal velocities of a water particle at the surface of the fluid and the body, respectively. 
Using Eqn.~(\ref{eq:detadt}) the velocity of a water particle is 
\begin{equation} \label{eq:Vw}
V_w = 
\pi f H
\sin(k x - 2\pi f t).
\end{equation}

\subsubsection{Inertia}

The inertial force, $F_{i}$, is 
\begin{equation} \label{eq:inertia}
F_i = m (1+C_m) \frac{dV}{dt}.
\end{equation}
Here, $C_m$ is the so-called added mass coefficient. 
It accounts for  `virtual mass'  due to fluid around the body. 

\subsection{Rumer Model} \label{sec:RumerMarchenko}

We derive the nonlinear ordinary differential equation (\textsc{ode}) 
\begin{equation*} \label{eq:ode}
m(1+C_m) 
\frac{d V}{ d t}= 
- m g
 \frac {\partial \eta} {\partial x} 
+ \rho_w 
C_d 
W
|V_w - V| 
(V_w - V),
\end{equation*}
for the velocity of the body, $V$, 
by combining Eqns.~(\ref{eq:sliding}),  (\ref{eq:drag}) and (\ref{eq:inertia}). 
Substituting Eqns. (\ref{eq:detadt}) and (\ref{eq:Vw}) into the \textsc{ode} 
 gives
\begin{multline*} \label{eq:ode2}
m(1+C_m) 
\frac{d V}{ d t} = 
 - (m g kH/2) \cos( k x - 2\pi f t ) \\
\ + \rho_w C_d W \left | {\pi f H \sin(k x - 2\pi f t) - V} \right | 
\\
\times \left ( \pi f H \sin(k x - 2\pi f t) - V \right ). 
\end{multline*}
The added mass and drag coefficients, $C_m$ and $C_d$, are, generally, determined experimentally, 
and are considered known.
By solving the above \textsc{ode}, we are able to predict the horizontal velocity of the body for an incident wave of prescribed height, wavenumber and frequency.

If we introduce the variable $\theta =k x - 2\pi f t$ the \textsc{ode} reduces to the autonomous system
\begin{subequations}
\begin{multline*} 
m(1+C_m) 
\frac{d V}{ d t} = 
 - (m g kH/2)  \cos( \theta)  
 \\
\ + \rho_w C_d W \left | { \pi f H \sin(\theta) - V} \right | \left ( \pi f H \sin(\theta) - V \right ), 
\end{multline*}
and
\begin{equation*}
\frac{d\theta}{d t} = k V - 2\pi f.
\end{equation*}
\end{subequations}
This system always has a stable periodic solution. 
Further, outside the very steep wave regime, the periodic solution is the only attractor in the system, i.e.\  all solutions tend to the periodic solution, 
regardless of the initial conditions.  
However, the period of the periodic solution does not exactly match the period of the wave, due to drift.  The amplitude in surge
is found by integrating the velocity over a single period. 


\section{Wavetank Experiments}

\begin{figure} 
\centering
\includegraphics[width=0.8\columnwidth]{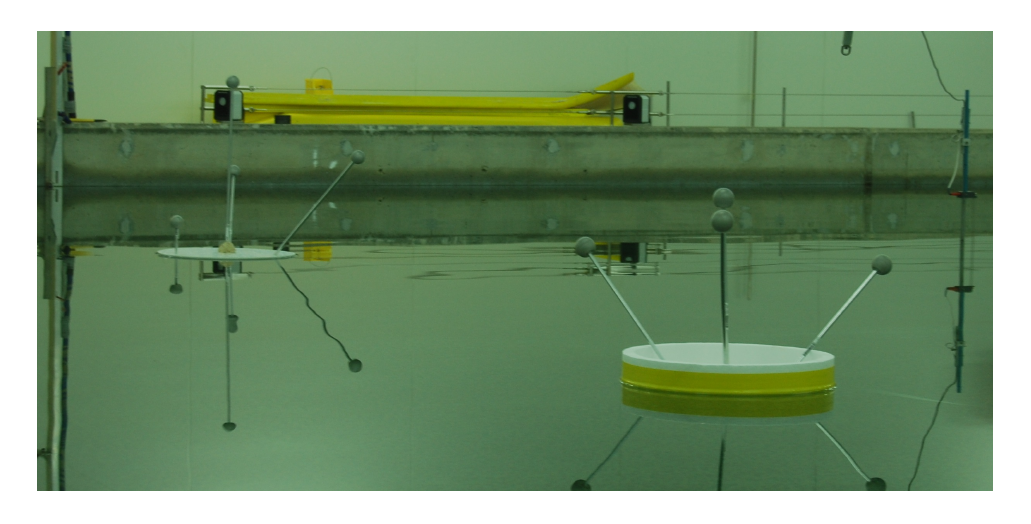}
\caption{Model floes. With edge barrier (foreground), and without (background).}
\label{fig:foto1}
\end{figure}

\begin{figure} 
\centering
\includegraphics[width=0.8\columnwidth]{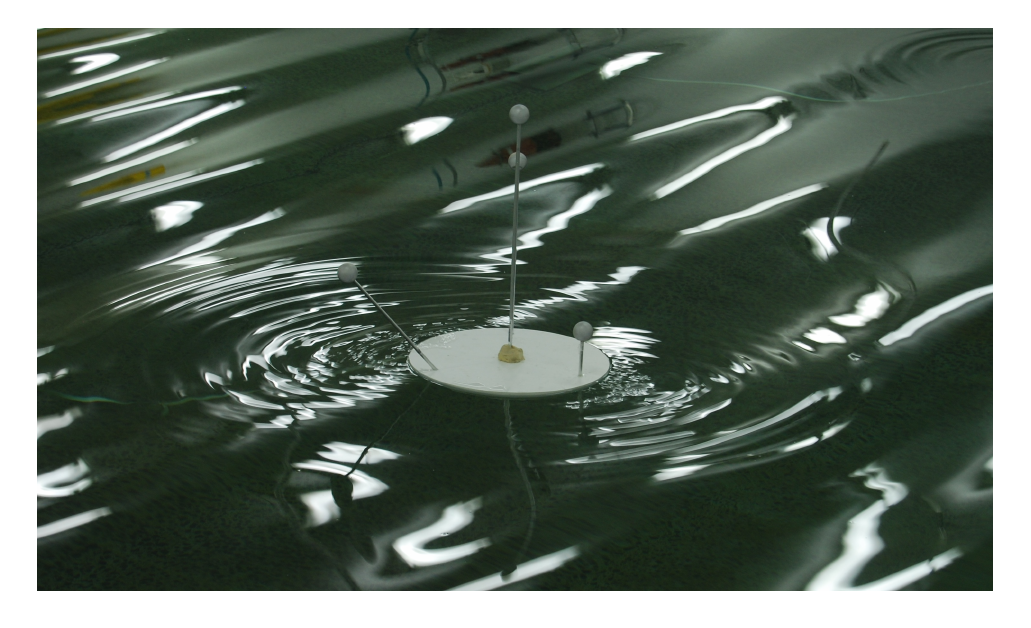}
\caption{Model floe during test. Scattered waves are evident.}
\label{fig:foto2}
\end{figure}

An experimental model of wave-induced surge motion of a floe was implemented in the model test basin (\textsc{mtb}) of the Australian Maritime College, Launceston, Tasmania.
The \textsc{mtb} is 35\,m long and 12\,m wide.

The floe was modelled by a thin Nycel$^{\text{\textregistered}}$ plastic disk, of 400\,mm diameter and 15\,mm thickness. 
At a scale of 1:100 this models an ice floe of 40\,m diameter and 1.5\,m thickness. 
Table~\ref{table:nonlin} provides a summary of parameters for the floe model.  

\begin{table}[ht]
\caption{Floe parameters}
\centering
\begin{tabular}{l @{\hspace{25pt}} l @{\hspace{25pt}} l }
\hline
\hline 
   & Model Scale & Full Scale\\
[0.5ex]
\hline
Material		&  Nycel	& 	Sea ice\\
Scale	         & 1	& 100 \\
Diameter           &  400\,mm	& 40\,m \\
Thickness	         & 15\,mm &	1.5\,m \\
Draft		         & 13\,mm		& 1.34\,m \\
Floe Mass		& 1.68\,kg		& 1.73 x 106\,kg \\
[1ex]
\hline
\end{tabular}
\label{table:nonlin}
\end{table}

Tests were conducted for the floe with and without a  polystyrene edge barrier. 
The barrier is 50\,mm high and 25\,mm thick.
It restricts overwash, i.e.\ the wave running over the top of the floe \citep{toffoli_ag}.
Fig.~\ref{fig:foto1} shows floes with and without an edge barrier.

Overwash is not yet included in theoretical models.
We therefore only present results from tests conducted with the barrier.
(Overwash has negligible impact on floe motions for large wavelengths, and
damps floe motions for smaller wavelengths.)

A mechanical wave maker was used to generate regular (monochromatic) waves.
Wave heights between 10\,mm and 80\,mm, and frequencies between 0.5\,Hz and 2\,Hz were tested, with corresponding wavelengths approximately 0.5\,m to 5\,m.
Wave gauges were located 1\,m in front of the equilibrium position of the floes, in order to verify the incident wave heights.

The ambient fluid depth was  800\,mm. 
The floes were moored to the \textsc{mtb} floor with a 1.6\,m long elastic band.
A static beach was located at the opposite end of the \textsc{mtb} to the wave maker, in order to reduce the waves reflected at this boundary.
The equilibrium position of the floes was 17.7\,m from the beach. 

Four spherical polystyrene markers, with 30\,mm diameters, mounted on light-weight aluminium cylinders, were attached to each floe.
The markers and mounts are visible in Figs.~\ref{fig:foto1} and \ref{fig:foto2}. 
The markers and mounts were sufficiently light to have negligible influence on the motions of the floes. 

The location of the markers was measured stereoscopically by the Qualysis$^{\text{\textregistered}}$ motion tracking system, via a set of infrared cameras. 
The Qualysis system provides time series of the translational  and rotational motions of the floes.
The translational motions provide the location of the floes with respect to a Cartesian coordinate system $(x,y,z)$.
The origin of the coordinate system is the equilibrium centre of the floe with a barrier. 
Horizontal translations are described by the $(x,y)$ coordinate, where $x$ denotes the location along the \textsc{mtb}, pointing from the wave maker to the beach, and $y$ denotes the location across the \textsc{mtb}.
Vertical motions, i.e. heave, are described by the $z$ coordinate.
The rotational motions provide the pitch, roll and yaw motions of the floes.

\subsection{Data Processing}

\begin{figure} 
\centering
\includegraphics[width=0.8\columnwidth]{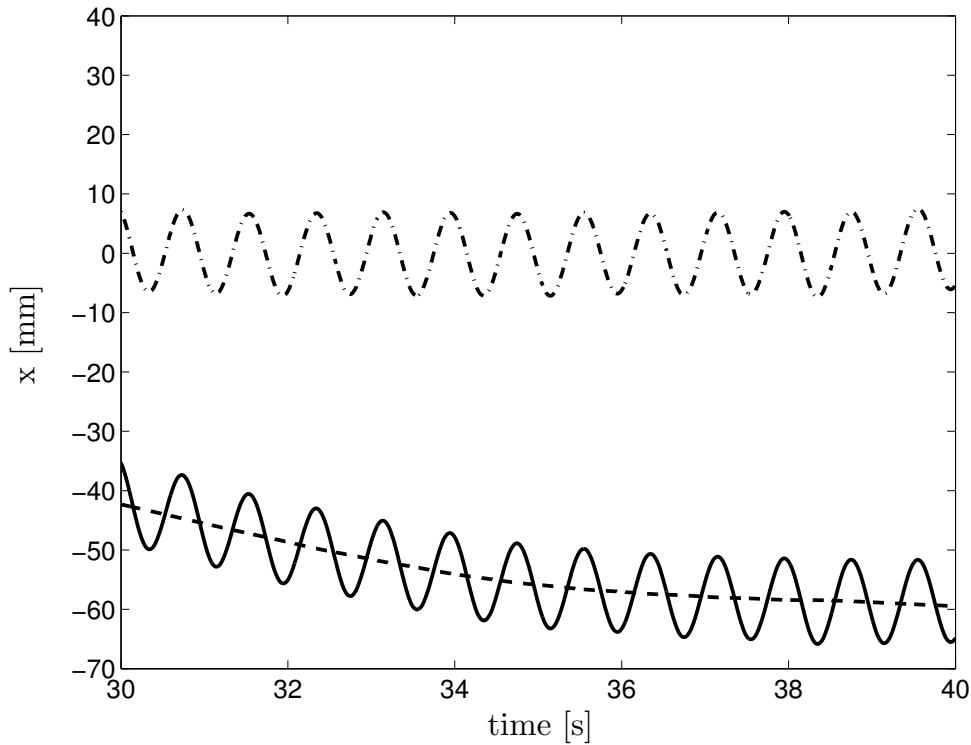}
\caption{Example decomposition of surge and drift motions. 
Full signal  (solid curve), drift (dashed) and surge (dot-dash).
}
\label{fig:DecomposedData}
\end{figure}

\begin{figure} 
\centering
\includegraphics[width=0.8\columnwidth]{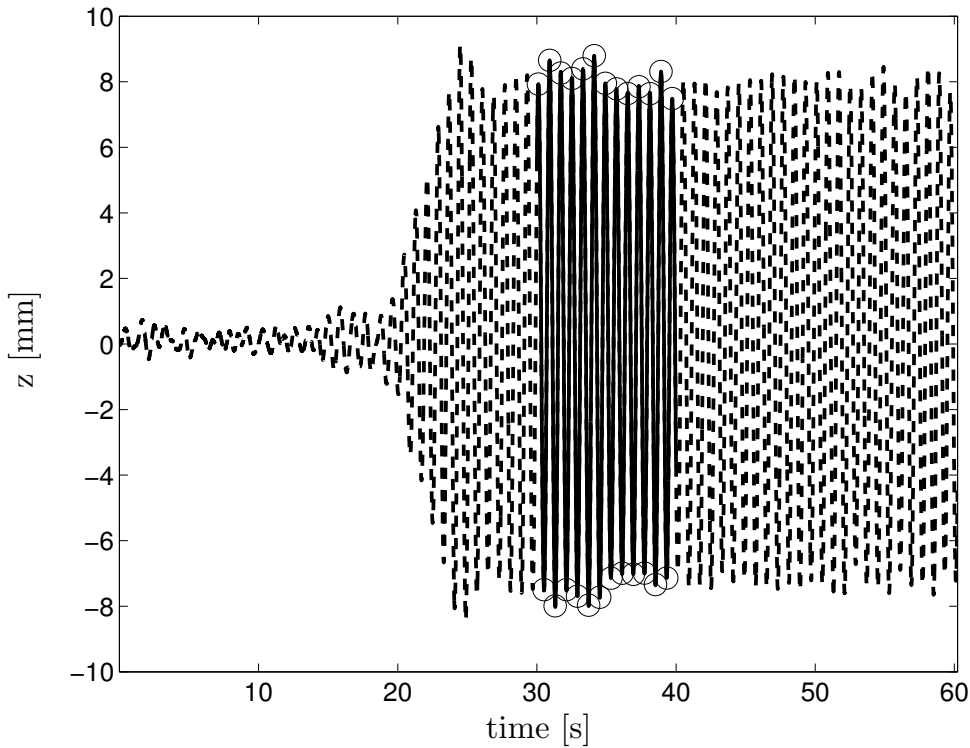}
\caption{An example of the steady-state window (solid curve).
Circles mark peaks and troughs within the window. }
\label{fig:SSData}
\end{figure}

Raw data is first smoothed to eliminate noise, 
via the \textsc{matlab} package function \texttt{smooth}. 
The \texttt{lowess} local regression method, using weighted linear least squares, 
is selected along with an appropriate smoothing factor.

The \texttt{smooth} function is also used to decouple the
surge and drift motions from the translational signal in the $x$-direction. 
For a sufficiently large smoothing factor, the \texttt{smooth} function eliminates the surge oscillations from the signal, and, thus, returns the drift.
The surge motion is then calculated as the difference between the signal and the drift.
An example of the decomposition method is shown in Fig.~\ref{fig:DecomposedData}.

During a given test, the magnitude of surge and drift motions vary with time due to, initially, transients in the first few wave fronts, and, later, 
interference caused by waves reflected by the beach. 
We therefore identify a steady-state window, i.e.\ a time interval beginning after incident wave transients have passed and finishing before reflected waves reach the floe, 
and analyse the signal in this interval only. 

The steady-state window can be calculated analytically using the wave celerity (phase speed).
In practice we use a conservative estimate of the steady-state window.
An example of the steady-state window is shown in Fig.~\ref{fig:SSData}.
The peaks and troughs of the signal within the steady-state window are also identified. 
Wave heights are calculated as the differences between consecutive peaks and toughs.
A representative amplitude for the motion in the steady state window, in this case the heave motion, 
is calculated as half the mean of the set of calculated heights.


\section{Results}

\begin{figure} 
\centering
\includegraphics[width=0.8\columnwidth]{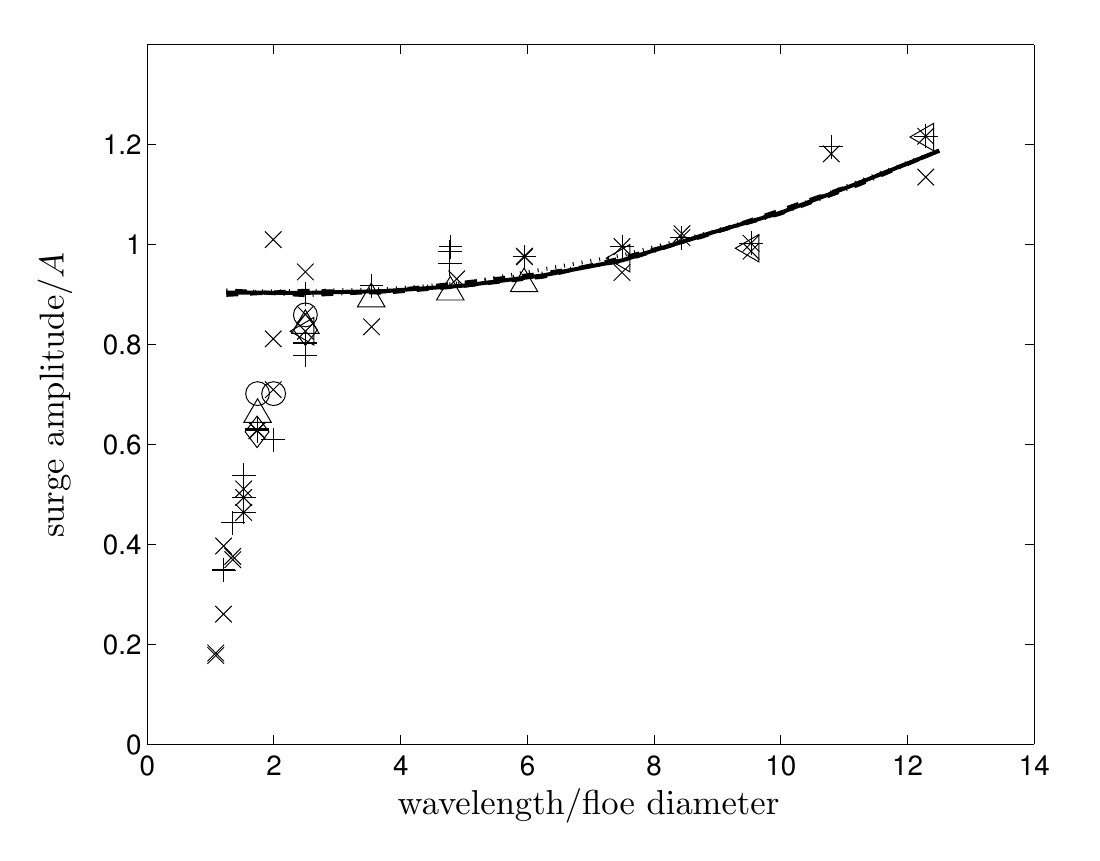}
\caption{Comparison of non-dimensional surge amplitude, as a function of non-dimensional wavelength, predicted by the Rumer model (curves) and experimental data (symbols). 
The Rumer model uses added mass $C_m = 0.1$, wave height $H = 10$\,mm, and  
drag coefficient $C_{d}=0$ (solid curve), 0.005 (dashed), 0.1 (dot-dash) and 0.5 (dotted).
Experimental data are taken from tests with wave heights $H=10$\,mm ($\circ$), 20\,mm ($\times$), 30\,mm ($+$), 40\,mm ($\ast$)
$H=50$\,mm ($\diamond$), 60\,mm ($\triangledown$), 70\,mm ($\triangle$), and 80\,mm ($\triangleleft$).
}
\label{Surge_Cm0p1H10}
\end{figure}

\begin{figure} 
\centering
\includegraphics[width=0.8\columnwidth]{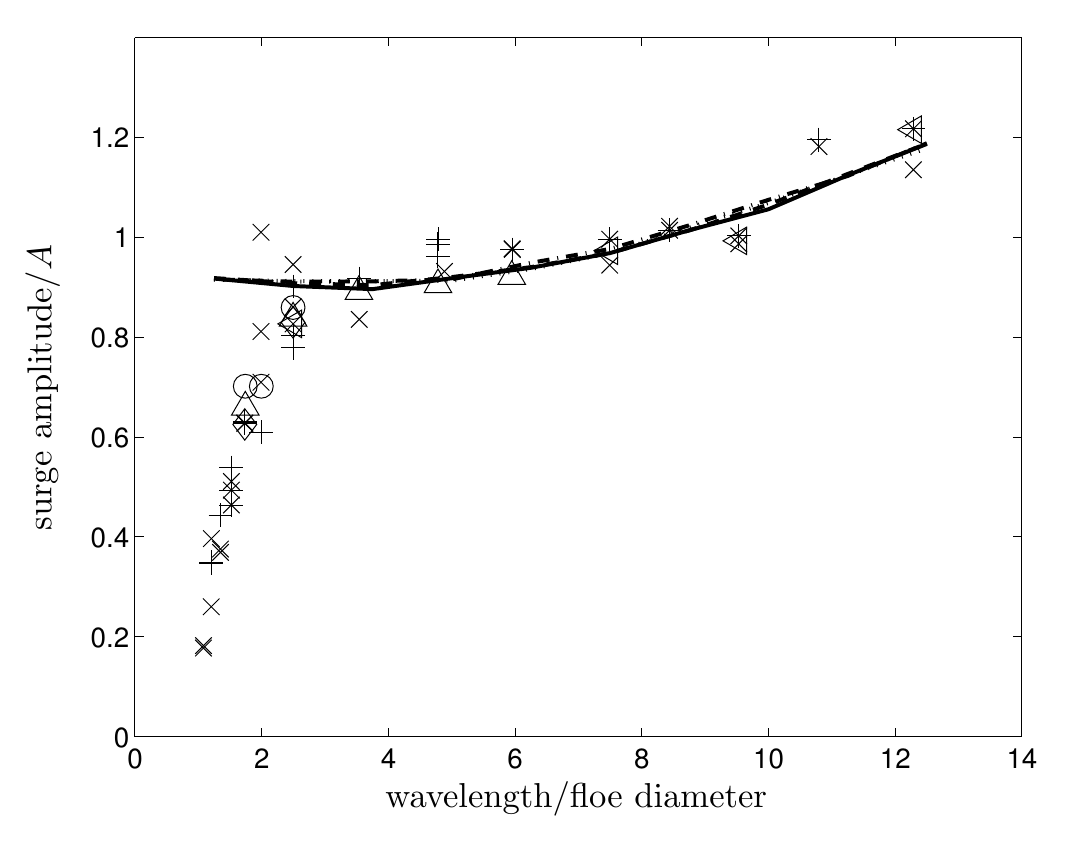}
\caption{As in Fig.~\ref{Surge_Cm0p1H10} but for $H = 40$\,mm. }
\label{Surge_Cm0p1H40}
\end{figure}

\begin{figure} 
\centering
\includegraphics[width=0.8\columnwidth]{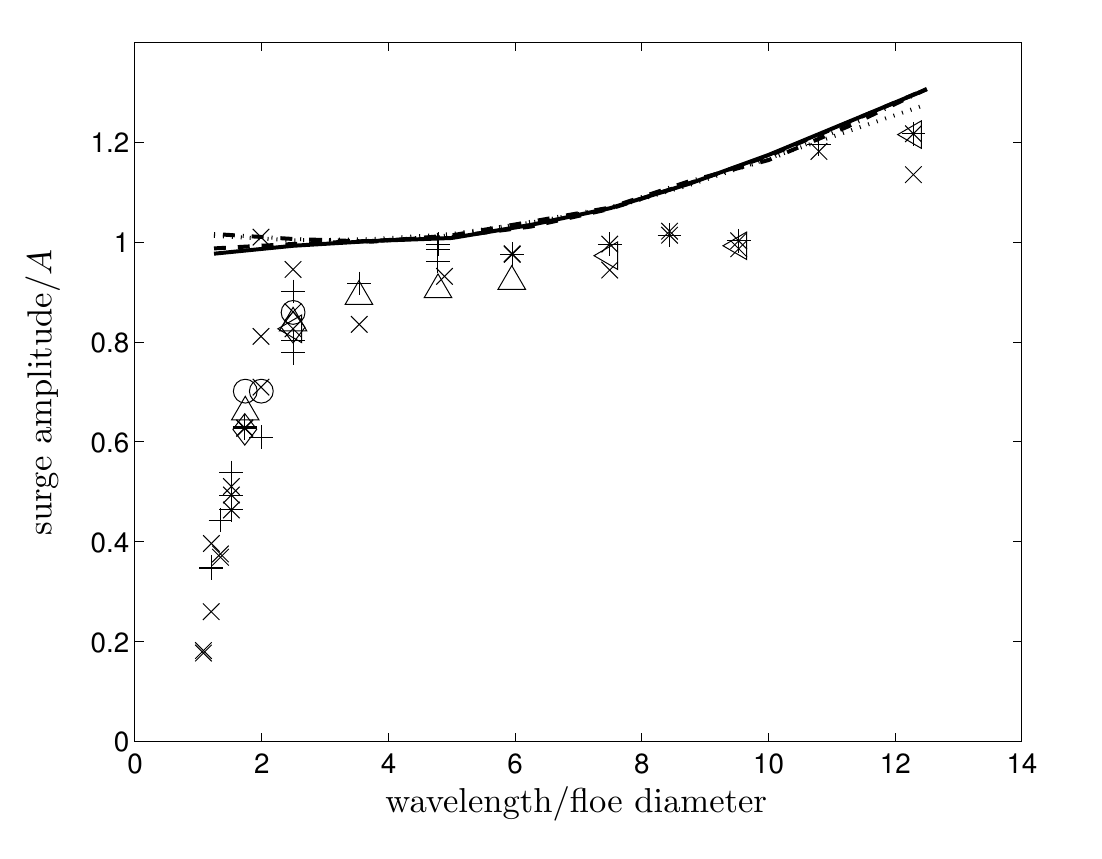}
\caption{As in Fig.~\ref{Surge_Cm0p1H10} but for $H = 40$\,mm and $C_m = 0$.  }
\label{Surge_Cm0H40}
\end{figure}

\begin{figure} 
\centering
\includegraphics[width=0.8\columnwidth]{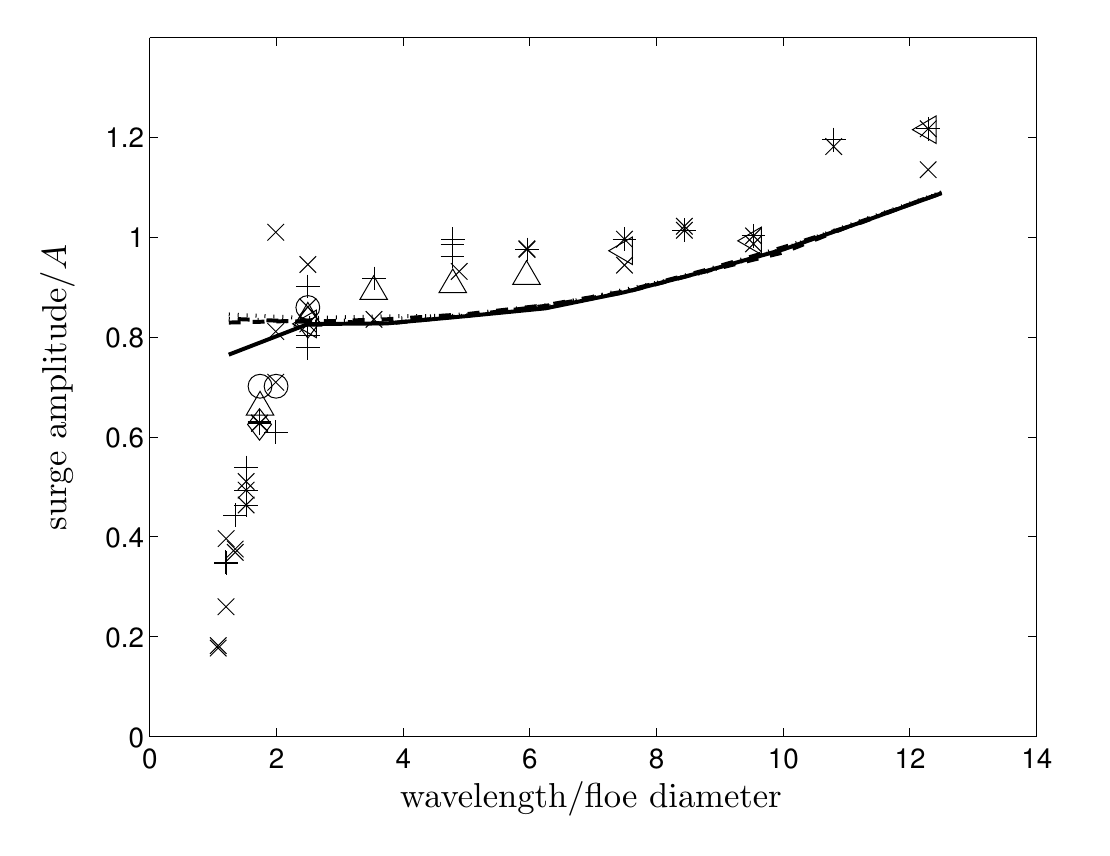}
\caption{As in Fig.~\ref{Surge_Cm0p1H10} but for $H = 40$\,mm and $C_m = 0.2$. }
\label{Surge_Cm0p2H40}
\end{figure}

\begin{figure} 
\centering
\includegraphics[width=0.8\columnwidth]{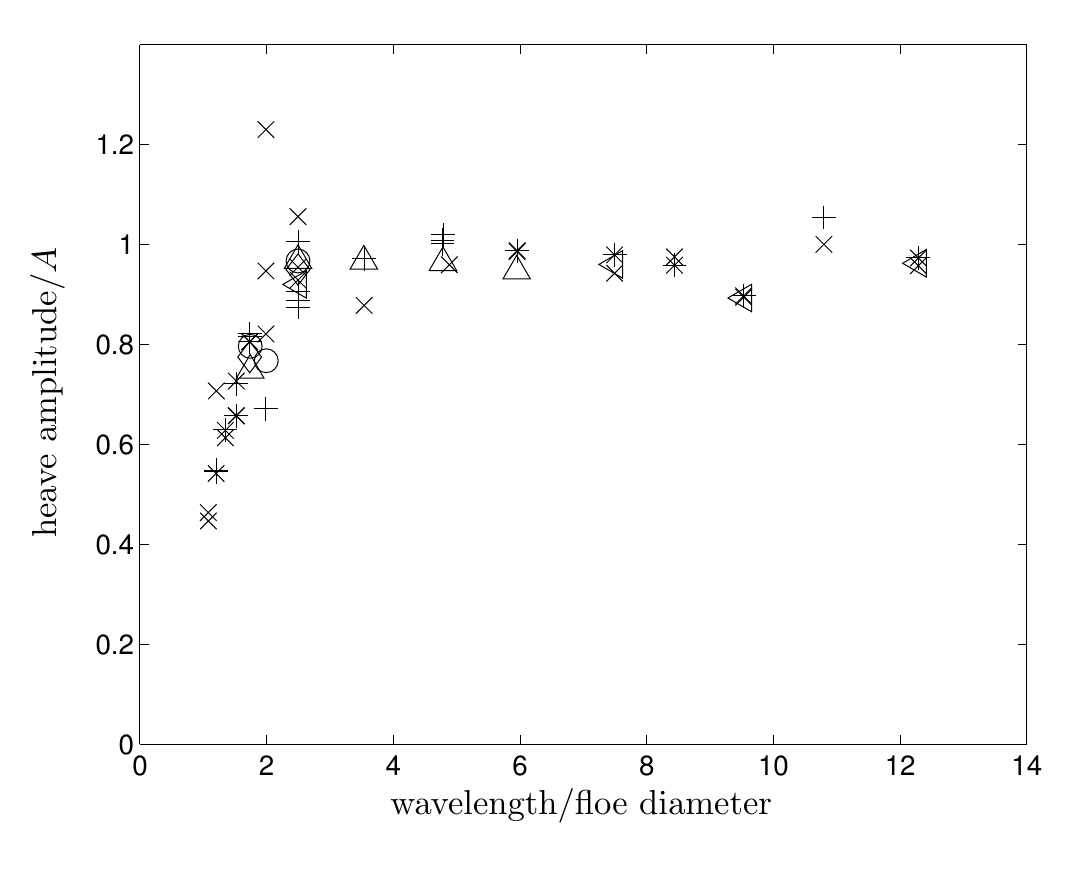}
\caption{Experimental data for heave motion. The results are labelled by approximate wave height as in  Fig.~\ref{Surge_Cm0p1H10}}
\label{fig:HeaveExp}
\end{figure}

\begin{figure} 
\centering
\includegraphics[width=0.8\columnwidth]{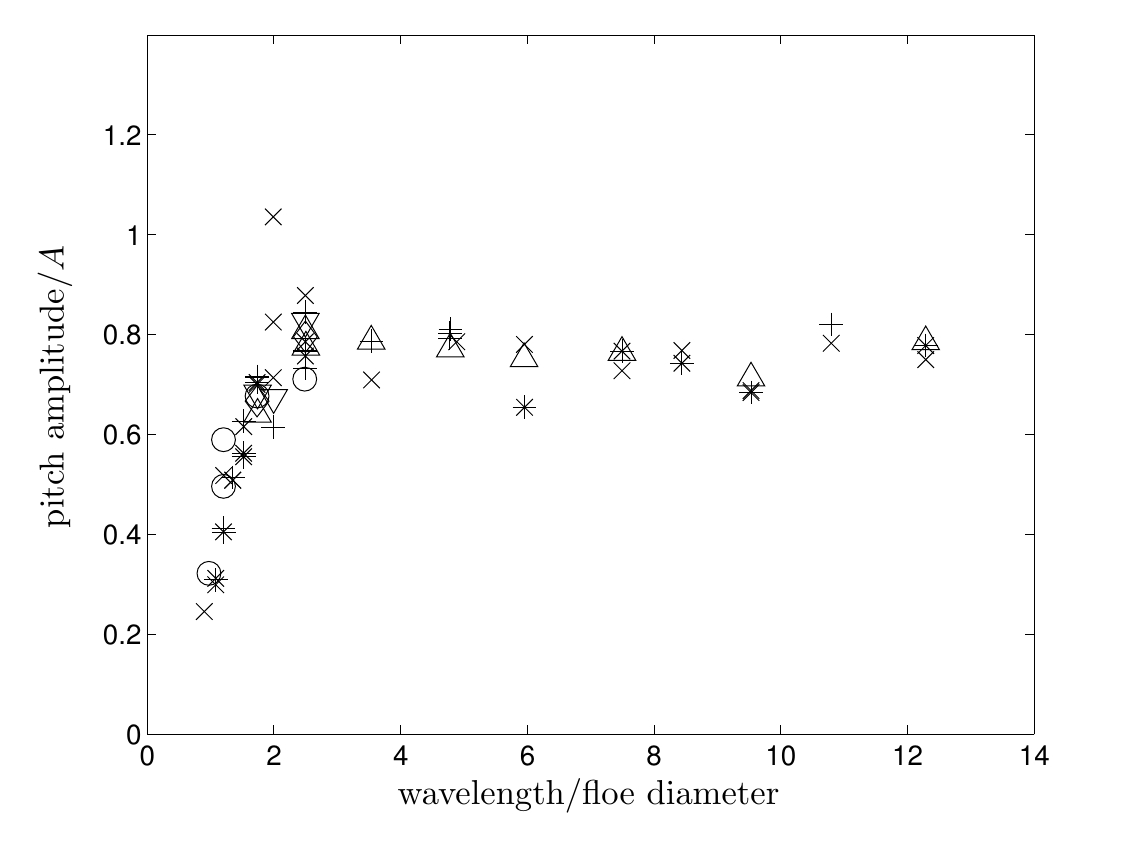}
\caption{Experimental data for pitch motion.  The results are labelled by approximate wave height as in Fig.~\ref{Surge_Cm0p1H10}}
\label{fig:PitchExp}
\end{figure}

Comparisons of surge amplitudes measured in the experiments and predicted by the Rumer model are shown in Figs.~\ref{Surge_Cm0p1H10} to \ref{Surge_Cm0p2H40}.
The surge amplitudes are non-dimensionalised with respect to the incident wave amplitudes, $A=H/2$. 
The wave heights used are those measured by the wave gauge in front of the floe.
Therefore, heights referred to from here on are the target heights. 
The non-dimensional surge amplitudes are presented as functions of wavelength, non-dimensionalised with respect to floe diameter.

The different figures show results for different added mass coefficients, $C_m$, and wave heights, $H$, in the Rumer model.
Four different values of the drag coefficient are considered for each added mass and wave height combination.


The experimental data are identical in the different figures.
The experimental data are grouped according to incident wave height. 
Very little spread is seen with respect to wave height. 
For the majority of the non-dimensional wavelengths considered, the experimental surge is approximately equal to the incident wave amplitude.
For large wavelengths,  the floe behaves like a particle on the surface.
As the particles perform elliptical orbits for finite fluid depths,
surge amplitudes are slightly larger than the incident amplitude for the largest wavelengths considered.

A regime change occurs when the wavelength is approximately equal to two floe diameters.
For wavelengths less than this transition length, 
the surge amplitude rapidly decreases with decreasing wavelength.
For the smallest wavelength considered, which is approximately equal to the floe diameter, the surge amplitude is approximately one-fifth of the incident amplitude.  



The surge amplitudes given by the Rumer model are insensitive to the value of the drag coefficient, $C_d$, for the majority of the non-dimensional wavelengths considered.
(If required, the drag coefficient can be measured experimentally by a towing test.)  
Surge is also insensitive to wave height.
Increasing/decreasing the added mass decreases/increases the surge motion, which can be understood intuitively.

Surge amplitudes predicted by the Rumer model agree with the experimental data 
for wavelengths approximately greater than two floe diameters --- the point at which a regime shift occurs in the experimental data.
The added mass coefficient $C_m=0.1$ provides the best agreement between the Rumer model and the experimental data (for wavelengths greater than two floe diameters). 



The non-dimensional heave and pitch motions measured in the experiments are shown in Figs.~\ref{fig:HeaveExp} and \ref{fig:PitchExp}, respectively.
(The Rumer model assumes that the floe follows the wave surface, i.e\ it assumes heave and pitch amplitudes equal to the incident wave amplitude.)  
A transition in the motions is again evident at a wavelength approximately equal to two floe diameters.



\section{Summary and discussion}

A theoretical and an experimental model of  surge motions of an ice floe, induced by regular waves, have been presented.
The theoretical model, the Rumer model, is based on a modified version of Morrison's equation.
The Rumer model assumes that the floe does not disturb the wave field.
Surge motions result from the force of gravity in conjunction with the slope of the waves, and drag force on the wetted surface of the floe.

The experimental model used a thin plastic disk to model the floe.
The model was implemented in a wave basin at scale 1:100.
The rigid-body motions of the floe were measured by a motion tracking system.
The raw time series data were processed to obtain the steady-state surge amplitude of the floe, 
for different wave heights and frequencies/wavelengths.

The surge amplitude, non-dimensionalised with respect to the incident wave amplitude, was presented as a function of wavelength, non-dimensionalised with respect to floe diameter.
The experimental data displayed a regime change from rapid increase in non-dimensional surge with increasing non-dimensional wavelength, to weak increase.
The transition between regimes occurred at a wavelength approximately twice as large as the floe diameter.
The regime change was also evident for heave and pitch motions in the experiments.

The Rumer model was able to  predict surge motions accurately for wavelengths larger than the transition wavelength.
An alternative theoretical model is clearly required for wavelengths less than two floe diameters.
In this regime, it is likely that a model based on linear potential flow theory will provide more accurate predictions of floe motions.
Potential flow theory includes modifications to the wave field surrounding the floe due to wave scattering. 


\section{Acknowledgements}
Experiments were funded by the Australian Maritime College. 
LB acknowledges funding support from the Australian Research Council (DE130101571). 
MM and LB acknowledge funding support from the Australian Antarctic Science Grant Program (Project 4123).

\end{document}